\documentclass[aps,pra,groupedaddress,showpacs,showkeys]{revtex4}
\usepackage{amssymb,bm}
\usepackage{graphicx}
\usepackage{amsmath}

\begin{document}

\title{Isoscalar amplitude dominance in  $e^+e-$ annihilation to  $N\bar{N}$ pair  close to the threshold. }

\author{V.F. Dmitriev}\email{V.F.Dmitriev@inp.nsk.su}
\author{A.I. Milstein}\email{A.I.Milstein@inp.nsk.su}
\author{S.G. Salnikov}\email{salsergey@gmail.com}
\affiliation{Budker Institute of Nuclear Physics, \\
and\\
Novosibirsk State University, 630090 Novosibirsk, Russia}

\date{\today}

\begin{abstract}
We use the Paris nucleon-antinucleon optical potential  for
explanation of experimental data in the process  $e^+e^- \rightarrow
p\bar p$ near  threshold.  It turns out that final-state interaction due to Paris
optical potential  allows us to  reproduce available experimental data.
It follows from  our consideration that  the isoscalar form factor is much larger than the isovector one.
\end{abstract}


\maketitle

\section{Introduction}
At present,  QCD can not describe  quantitatively the low-energy
nucleon-antinucleon interaction, and various  phenomenological
approaches have been suggested in order to explain numerous
experimental data , see , e.g., Refs.
\cite{pot82,bonn91,bonn95,paris82,paris94,Partial94} and recent
reviews \cite{KBMR02,KBR05}. However, parameters of the models still
can not be extracted with a good accuracy  from the experimental
data \cite{Rich95}.

Very recently, renewed  interest in low-energy nucleon-antinucleon
physics has been stimulated by the experimental observation of a
strong enhancement of decay probability at low invariant mass of
$p\bar p$ in the processes  $J/\Psi\to \gamma p\bar p$ \cite{BES03},
$B^+\to K^+p\bar p$ and $B^0\to D^0 p\bar p$
\cite{Belle02a,Belle02b,BaBar05}, $B^+\to \pi^+ p\bar p$ and $B^+\to
K^0 p\bar p$ \cite{Belle04c}, $\Upsilon\to \gamma p\bar p$
\cite{CLEO06}. One of the most natural explanation of this
enhancement is final state interaction of the proton and antiproton
\cite{Ker04,Bugg04,Zou04,Loi05,Sib05,Sib06}.

A similar phenomenon was observed  in the investigation of the
proton (antiproton) electric, $G_E(Q^2)$, and  magnetic, $G_M(Q^2)$,
form factors  in the process $e^+e^- \to p\bar p$
\cite{Bardin94,Armstrong93,Aubert06}. Namely, it was found that the
ratio $|G_E(Q^2)/G_M(Q^2)|$ strongly depends on $Q^2=4E^2$ (in the
center-of-mass frame) in the narrow  region of  the energy $E$  near
the  threshold of $p\bar p$ production. Such  strong dependence at
small $E$ is related to the large-scale interaction of proton and
antiproton. Therefore, it is possible to apply the approaches of
\cite{pot82,bonn91,bonn95,paris82,paris94,Partial94} for an
explanation of experimental data in the process  $e^+e^- \rightarrow
p\bar p$. In the present paper, we use the Paris nucleon-antinucleon
optical potential $V_{N\bar N}$  which has the form
\cite{paris82,paris94}:
\begin{equation}\label{VNN}
V_{N\bar N} = U_{N\bar N}-i\, W_{N\bar N} \, ,
\end{equation}
where the real part $U_{N\bar N}$  is the  $G$-parity transform of
the well established Paris $NN$ potential for the long- and
medium-range distances ($r\gtrsim 1\mbox{fm}$), and some
phenomenological part for the short distances. The absorptive part
$W_{N\bar N}$ of the optical potential takes into account the
inelastic channels of $N\bar N$ interaction, i.e. annihilation into
mesons. It is essential at short distances and depends on the
kinetic energy of the particles. We perform calculations in the non-relativistic approximation.
The Coulomb interaction between proton and antiproton is important only for the kinetic energy  $T\lesssim (\pi\alpha)^2 M \sim 1$~MeV, where $\alpha$ is the fine structure constant, and $M$ is the proton mass. Here we consider the process for kinetic energies $T\gg 1$~MeV, therefore we neglect the Coulomb interaction.  
 
Taking into account that the difference of the cross sections  $e^+e^- \rightarrow
p\bar p$ and  $e^+e^- \rightarrow n\bar n$ is small \cite{Aubert06,Antonelli98},  we calculate  the cross sections at a given isospin final states and   compare them  with the
 experimental data for $e^+e^- \rightarrow p\bar p$. As a result we found that the amplitude with the isospin $I=0$  strongly dominates.
  Besides, our prediction for the ratio  $|G_E(Q^2)/G_M(Q^2)|$ depends on the parameters of the Paris potential but independent 
  of the form factor at threshold. The ratio is also in a qualitative agreement with experimental data.

\section{Amplitude of the process}
In the nonrelativistic  approximation, the amplitude of $N\bar{N}$
pair production in a certain isospin channel $I=0,\,1$ near  threshold can be presented as follows (in units $4\pi\alpha/Q^2$):
\begin{eqnarray}\label{2}
&&T_{\lambda\mu}^I=\sqrt{2} \bm{\epsilon}_\lambda^* \Bigg[\mathcal
G_s^I\bm e_\mu +\mathcal G_d^I \frac{{\bm k}^2{\bm e}_\mu-3({\bm k}\cdot{\bm e}_\mu){\bm k}}{6M^2}\Bigg]\, ,\nonumber\\
&&\mathcal G_s^I=\mathcal F_1^I(Q^2)+\mathcal F_2^I(Q^2)+\frac{\beta^2}{6}[\mathcal F_2^I(Q^2)-\mathcal F_1^I(Q^2)]\,,\nonumber\\
&&\mathcal G_d^I=\mathcal F_1^I(Q^2)-\mathcal F_2^I(Q^2)\,,
\end{eqnarray}
where $\beta=k/M\ll 1$ , ${\bf e}_\mu$ is a virtual photon polarization vector, corresponding to the projection
of spin  $J_z=\mu=\pm 1$,  and $\bm{\epsilon}_\lambda$ is the spin-1 function of $N\bar{N}$ pair, $\lambda=\pm1,\, 0$ is the projection of spin on the vector $\bm k$. Two tensor structures in Eq.(\ref{2}) correspond to the s-wave and
d-wave production amplitudes. The total angular momentum of the
$N\bar{N}$ pair is fixed by a production mechanism. The functions
$\mathcal F_1^I(Q^2)$ and $\mathcal F_2^I(Q^2)$ are the Dirac form factors of the $N\bar{N}$ pair which include the  effects of final state interaction and have  a pronounced $Q^2$ behavior near the threshold. Summation over the polarization of  nucleon  pair and averaging over the  polarization of   virtual photon is performed using the equations,
\begin{equation}
\sum_{\lambda=1,2,3} \epsilon_\lambda^{i*}\epsilon_\lambda^j=\delta^{ij}\,,\quad
 \frac{1}{2}\sum_{\mu=1,2} e_\mu^{i*}e_\mu^j=\frac{1}{2}\delta^{ij}_\perp=\frac{1}{2}(\delta^{ij}-P^iP^j/P^2)\,,
\end{equation}
where $\bm P$ is the electron  momentum.

Our aim is to single out the  effects of final state interaction. In order to do that, we write the amplitude (\ref{2}) in the form
\begin{eqnarray}\label{3}
T_{\lambda\mu}^I & =&\sqrt{2} \int \frac{d^3p}{(2\pi)^3}{\bm\Phi}^{I(-)*}_{\bm k\lambda}({\bm p})\cdot \Bigg[G_s^I\bm e_\mu +G_d^I\frac{{\bm p}^2{\bm e}_\mu-3({\bm p}\cdot{\bf
e}_\mu){\bm p}}{6M^2}\Bigg]\, ,
\end{eqnarray}
where $\bm\Phi^{I(-)}_{\bm k\lambda}({\bm p})$ is the Fourier transform of the function $\bm\Psi^{I(-)}_{\bm k\lambda}({\bm r})$, the wave function of the $N\bar{N}$ pair in  coordinate space.
This wave function is the  solution of the Schr\"odinger equation
\begin{equation}
\mathbf{\Psi}^{I(-)*}_{{\bm k}\lambda}({\bm r})\hat{H}=M\beta^2
\mathbf{\Psi}^{I(-)*}_{{\bm k}\lambda}({\bm r})\,,\quad
\hat{H}=\frac{{\bm p}^2}{M}+V_{N\bar{N}}\,,
\end{equation}
where $V_{N\bar{N}}$ is the optical potential. Note that
$\mathbf{\Psi}^{I(-)*}_{{\bm k}\lambda}({\bm r})$ is the left
eigenfunction of the bi-orthogonal set of eigenfunctions of the
non-Hermitian operator $\hat{H}$. The asymptotic form of the wave function at large distances reads
\begin{equation}
\bm\Psi^{I(-)}_{\bm k\lambda}({\bm r})\approx \bm{\epsilon}_\lambda\,e^{i\bm k\cdot\bm r}+f_{\lambda\lambda'} \frac{e^{-ikr}}{r}\bm{\epsilon}_{\lambda'} \,.
\end{equation}
In Eq.(\ref{3}) the form factors $G_s$ and $G_d$  are
\begin{eqnarray}\label{31}
&&G_s^I=F_1^I+ F_2^I+\frac{\beta^2}{6}(F_2^I-F_1^I)\,,\nonumber\\
&& G_d^I=F_1^I- F_2^I\,,
\end{eqnarray}
where the "bare" Dirac form factors  $F_1$ and  $F_2$  do not account for the effect of
final state interaction. Near threshold  these form factors  are  smooth
functions of $Q^2$ and can be treated  as  phenomenological constants.

\section{Wave function}

Let us introduce the vector spherical functions $\bm Y_{J\,\mu}^L(\bm n)$ as
\begin{eqnarray}\label{A}
\bm Y_{J\,\mu}^L(\bm n)=\sum_m  C^{J\,\mu}_{L\,m,\,1\,\mu-m}\,Y_{L\,m}(\bm n) \bm \epsilon_{\mu-m}\,,
\end{eqnarray}
where $Y_{L\,m}(\bm n)$ are spherical harmonics,  $C^{J\mu}_{L\,m,\,1\,\mu-m}$ are
Clebsch-Gordan coefficients, and $\bm n=\bm r/r$. In Eq. (\ref{A}) the quantization axes is directed along the vector $\bm k$. Then the wave function $\bm\Psi^{I(-)}_{\bm k\lambda}({\bm r})$ can be written in the form
\begin{eqnarray}\label{WFpsi}
&&\bm\Psi^{I(-)}_{\bm k\lambda}({\bm r})=\sum_J \sqrt{4\pi(2J+1)} C^{J\,\lambda}_{J\,0,\,1\,\lambda} v^{I*}_J(r)\bm Y_{J\,\lambda}^J(\bm n)\nonumber\\
&&+\sum_J \sqrt{4\pi(2J-1)} C^{J\,\lambda}_{J-1\,0,\,1\,\lambda}\Big[ u^{I*}_{1J}(r)\bm Y_{J\,\lambda}^{J-1}(\bm n)+w^{I*}_{1J}(r)\bm Y_{J\,\lambda}^{J+1}(\bm n)\Big]                              \nonumber\\
&&+\sum_J \sqrt{4\pi(2J+3)} C^{J\,\lambda}_{J+1\,0,\,1\,\lambda}\Big[ u^{I*}_{2J}(r)\bm Y_{J\,\lambda}^{J-1}(\bm n)+w^{I*}_{2J}(r)\bm Y_{J\,\lambda}^{J+1}(\bm n)\Big]\,,                               \end{eqnarray}
 Here the functions $v^I_J(r)$,  $u^I_{nJ}(r)$, and $w^I_{nJ}(r)$ have the asymptotic form at large distances
\begin{eqnarray}\label{asvuw}
&&v^I_J(r)=\frac{1}{2ikr}\Big[S_0^{IJ}e^{i[kr-J\pi/2]}-e^{-i[kr-J\pi/2]}\Big]\,,\nonumber\\
&&u^I_{1J}(r)=\frac{1}{2ikr}\Big[S_{11}^{IJ}e^{i[kr-(J-1)\pi/2]}-e^{-i[kr-(J-1)\pi/2]}\Big]\,,\nonumber\\
&&w^I_{1J}(r)=\frac{1}{2ikr}S_{12}^{IJ}e^{i[kr-(J+1)\pi/2]}\,,\nonumber\\
&&u^I_{2J}(r)=\frac{1}{2ikr}S_{21}^{IJ}e^{i[kr-(J-1)\pi/2]}\,,\nonumber\\
&&w^I_{2J}(r)=\frac{1}{2ikr}\Big[S_{22}^{IJ}e^{i[kr-(J+1)\pi/2]}-e^{-i[kr-(J+1)\pi/2]}\Big]\,,
\end{eqnarray}
where $S_0^{IJ}$  and $S_{ij}^{IJ}$ are some functions of energy with $|S_0^{IJ}|\leqslant 1$ and $|S_{ij}^{IJ}|\leqslant 1$. Due to angular momentum conservation, only the terms with $J=1$ and $L=J\pm 1$ (i.e., $L=0,\,2$) contribute to the matrix element (\ref{3}). Then the amplitude (\ref{3}) can be written as
\begin{eqnarray}\label{dress1}
T_{\lambda\mu}^I & =&\sqrt{2}\lim_{r\rightarrow 0}\Bigg[G_s^I\bm e_\mu  -G_d^I\frac{{\bm e}_\mu{\triangle}-3({\bm \nabla}\cdot{\bf e}_\mu){\bm \nabla}}{6M^2}\Bigg]\mathbf{\psi}^{I*}_{{\bm k}\lambda}(\bm r)\, ,\nonumber\\
&&\psi^{I}_{{\bm k}\lambda}(\bm r)= [u^{I*}_{11}(r)\bm \epsilon_\lambda+w^{I*}_{11}(r)\sqrt{4\pi}\bm Y_{1\,\lambda}^2(\bm n)]\nonumber\\
&&+ \sqrt{5}\, C^{1\,\lambda}_{2\,0,\,1\,\lambda}[u^{I*}_{21}(r)\bm \epsilon_\lambda+w^{I*}_{21}(r)\sqrt{4\pi}\bm Y_{1\,\lambda}^2(\bm n)]\,.
\end{eqnarray}
Finally we have
\begin{eqnarray}\label{21}
&&\mathcal G_s^I= G_s^I\,u^{I}_{11}(0)+\frac{5\, G_d^I }{\sqrt{2}M^2}\lim_{r\rightarrow 0}\left(\frac{w^{I}_{11}(r)}{r^2}\right)\,,\nonumber\\
&&\mathcal G_d^I=\frac{6G_s^I}{\sqrt{2}\beta^2}\,u^{I}_{21}(0)+15\, G_d^I \lim_{r\rightarrow 0}\left(\frac{w^{I}_{21}(r)}{k^2r^2}\right)\,.
\end{eqnarray}
The first term in $\mathcal G_d^I$ contains the large factor $6/\beta^2$ while the second term in $\mathcal G_s^I$ is small due to the proton mass $M$ in denominator. Thus, in the non-relativistic approximation the amplitude $T_{\lambda\mu}^I$ reads,
\begin{eqnarray}\label{dress2}
T_{\lambda\mu}^I & =&G_s^I\Bigg\{ \sqrt{2}u^{I}_{11}(0)(\bm e_\mu\cdot \bm{\epsilon}_\lambda^*)+
 u^{I}_{21}(0)[ ({\bm e}_\mu\cdot \bm{\epsilon}_\lambda^*)-3(\hat{\bm k}\cdot  {\bm e}_\mu)(\hat {\bm k}\cdot\bm{\epsilon}_\lambda^*)]\Bigg\}\, \,,
\end{eqnarray}
where $\hat{\bm k}=\bm k/k$. The interpretation of this equation is the following.  As a result of re-scattering due to the tensor forces, the pair  produced  at the origin in $s$-wave has a non-zero amplitude to transfer to $d$-wave.

\section{Cross section and Sachs form factors}

The cross section corresponding to the amplitude (\ref{3}) has the form in the center-of-mass frame (see, e.g., Ref.~\cite{BLP1971})
\begin{eqnarray}\label{CS0}
\frac{d\sigma}{d\Omega} & =&\frac{\beta\alpha^2}{4Q^2}\,\Bigg[|G_M(Q^2)|^2(1+\cos^2\theta)+\frac{4M^2}{Q^2}\,
|G_E(Q^2)|^2\sin^2\theta\Bigg]\,.
\end{eqnarray}
Here $\theta$ is the angle between the electron (positron) momentum $\bm P$ and the momentum of the final particle $\bm k$. In terms of the "dressed" form factors ${\cal G}_s^I$ and ${\cal G}_d^I$ the electromagnetic Sachs form factors, corresponding to the contribution of the amplitude with the isospin $I$,  have the form
\begin{eqnarray}\label{FFnr}
&&G_M^I={\cal G}_s^I+\frac{\beta^2}{6}{\cal G}_d^I=G_s^I[u^{I}_{11}(0)+\frac{1}{\sqrt{2}}u^{I}_{21}(0)]
\,,\nonumber\\
&& \frac{2M}{Q}G_E^I={\cal G}_s^I-\frac{\beta^2}{3}{\cal G}_d^I=G_s^I[u^{I}_{11}(0)-{\sqrt{2}}u^{I}_{21}(0)] \,.
\end{eqnarray}
Thus, in the non-relativistic approximation the ratio $G_E^I/G_M^I$ is independent of the constant $G_s^I$,
\begin{eqnarray}\label{GEGMnr}
&&\frac{G_E^I}{G_M^I}=
\dfrac{u^{I}_{11}(0)-{\sqrt{2}}u^{I}_{21}(0)}{u^{I}_{11}(0)+\dfrac{1}{\sqrt{2}}u^{I}_{21}(0)}
\,.
\end{eqnarray}
Note that the electromagnetic interaction is important only in the narrow region where $\beta\sim \pi\alpha$ and the nucleon energy is $E=M\beta^2/2\sim 0.3$MeV. In this paper we will not consider this narrow region and neglect the electromagnetic interaction in the potential. Then,
the amplitude of $p\bar{p}$ pair production, $T^{(p)}_{\lambda\mu}$,  and the amplitude of $n\bar{n}$ pair production,  $T^{(n)}_{\lambda\mu}$, has the form
$$T^{(p)}_{\lambda\mu}=\frac{T^{1}_{\lambda\mu}+T^{0}_{\lambda\mu}}{\sqrt{2}}\,,\quad
T^{(n)}_{\lambda\mu}=\frac{T^{1}_{\lambda\mu}-T^{0}_{\lambda\mu}}{\sqrt{2}}\,.$$
The contribution of the isospin $I$ to the total cross section of the nucleon pair production reads
\begin{eqnarray}\label{CStot}
\sigma^I & =&\frac{2\pi\beta\alpha^2}{Q^2}\,|G_s^I|^2\,[|u^I_{11}(0)|^2+|u^I_{21}(0)|^2]\,.
\end{eqnarray}
Thus, to describe the energy dependence of the ratio ${G_E^I}/{G_M^I}$ and the cross section
$\sigma^I$ in the non-relativistic approximation, it is necessary to know the functions
$u^I_{11}(0)$ and $u^I_{21}(0)$.

Let us write the hamiltonian  $H^I$ for the isospin $I$ as follows,

\begin{eqnarray}
&&H^I=\frac{p_r^2}{M}+V_0^I(r)\delta_{L0}+V_2^I(r)\delta_{L2}+V_3(r)\,S_{12}\,,\nonumber\\
&&S_{12}=6(\bm S\cdot\bm n)^2-4\,,
\end{eqnarray}
where $\bm S$ is the spin operator for the  spin-one system of produced pair,  $(-p_r^2)$ is the radial part of the Laplace operator, and $L$ denotes the orbital angular momentum.  Then the radial wave functions $u^I_{n1}$ and $w^I_{n1}$ , $n=1,\, 2$~, satisfy  the equations

\begin{eqnarray}\label{equation}
&&\frac{p_r^2}{M}\chi +{\cal V}\chi=2E\chi\,,\nonumber\\
&&{\cal V}=\begin{pmatrix}V_0^I & -2\sqrt{2}\,V_3^I \\
-2\sqrt{2}\,V_3^I  & V_2^I-2V_3^I
\end{pmatrix}\,,\quad \chi=\begin{pmatrix}u^{I}_{n1}\\w^{I}_{n1}\end{pmatrix}\,.
\end{eqnarray}
The asymptotic form of the solutions at large distances is, Eq. (\ref{asvuw}),
\begin{eqnarray}\label{asvuw1}
&&u^{I}_{11}(r)=\frac{1}{2ikr}\Big[S_{11}^{I1}\,e^{ikr}-e^{-ikr}\Big]\,,\nonumber\\
&&w^{I}_{11}(r)=-\frac{1}{2ikr}S_{12}^{I1}\,e^{ikr}\,,\nonumber\\
&&u^{I}_{21}(r)=\frac{1}{2ikr}S_{21}^{I1}\,e^{ikr}\,,\nonumber\\
&&w^{I}_{21}(r)=\frac{1}{2ikr}\Big[-S_{22}^{I1}e^{ikr}+e^{-ikr}\Big]\,.
\end{eqnarray}

\section{Results and discussion}

It is known that the difference of the cross section of the processes  $e^+e^- \rightarrow
p\bar p$ and  $e^+e^- \rightarrow n\bar n$ is small \cite{Aubert06,Antonelli98}. Therefore it is natural to suggest that one of the isospin amplitudes is much larger than  another one.
Fig.~1 shows the cross section $\sigma^0$ at $| G_s^0|^2=101.8$ (solid line) as well as the cross section  $e^+e^- \rightarrow p\bar p$. The only free parameter $| G_s^0|^2$ has been found by normalizing the theoretical curve to the data at the third experimental point above threshold.  
It is seen that the cross section for $I=0$ perfectly reproduces the shape of the data in a wide range of energy near threshold (from threshold up to $2E=2.2$~GeV). Besides, for this case   
the value of $| G_s^0|=|F_1^0+F_2^0|\approx 10$ looks reasonable compared to the value $ |G_s^1|\approx 100$ obtained by assumption of isovector dominance.  

Fig.~2 shows the cross section $\sigma^1$  normalized at the same data point. It is seen that the shape of the theoretical curve does not reproduce the energy behaviour of the measured cross section. In addition, the value of the fitting parameter $ |G_s^1|^2=10184$ does not look reasonable.

Fig.~3 shows the ratio $|{G_E^0}/{G_M^0}|$, see Eq.(\ref{GEGMnr}), together with the experimental data Ref.\cite{Aubert06}. Our prediction is in a qualitative agreement with the data.

Isoscalar dominance in a model of final-state interaction based on the Paris potential has a simple explanation. At very small distances, the Paris potential corresponding to $I=1$ is strongly repulsive while the potential for  $I=0$ is strongly  attractive. As a result, the wave function for $I=1$ at small distances is strongly suppressed as compare to  the case   $I=0$. 

To conclude, using the Paris optical potential, we show that the isoscalar amplitude dominates in the cross section of the process   $e^+e^- \rightarrow N\bar N$. Our prediction for the cross section is in good agreement with the data. The prediction for  $|{G_E^0}/{G_M^0}|$ obtained under assumption of the isoscalar dominance agrees qualitatively with the data. To confirm our statement on the isoscalar dominance, other optical potential models should be considered. 

\section*{Acknowledgements}
The work  was supported by the Ministry of Education and Science of the
Russian Federation.

\newpage
%
\begin{figure}
\includegraphics{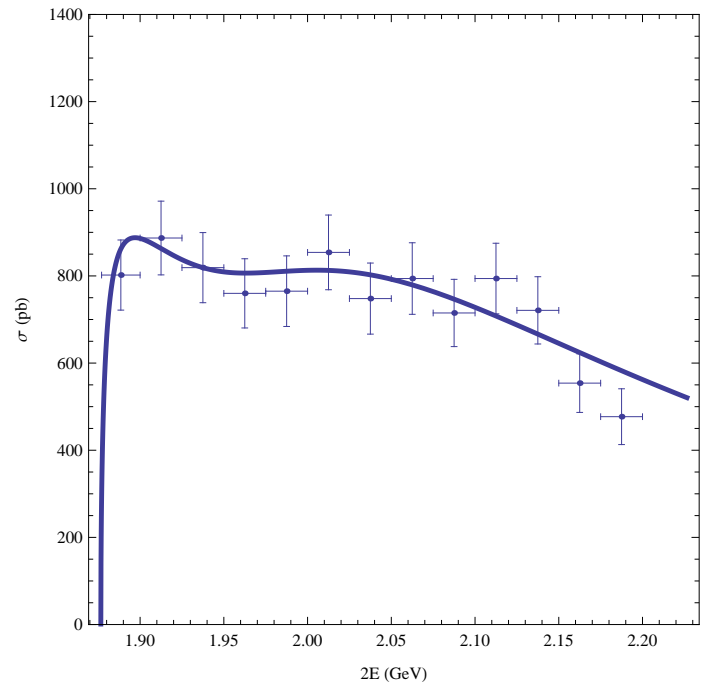}
\caption{Calculated isoscalar cross section, normalized to the data at the third point. }
\end{figure}
\newpage
%
\begin{figure}
\includegraphics{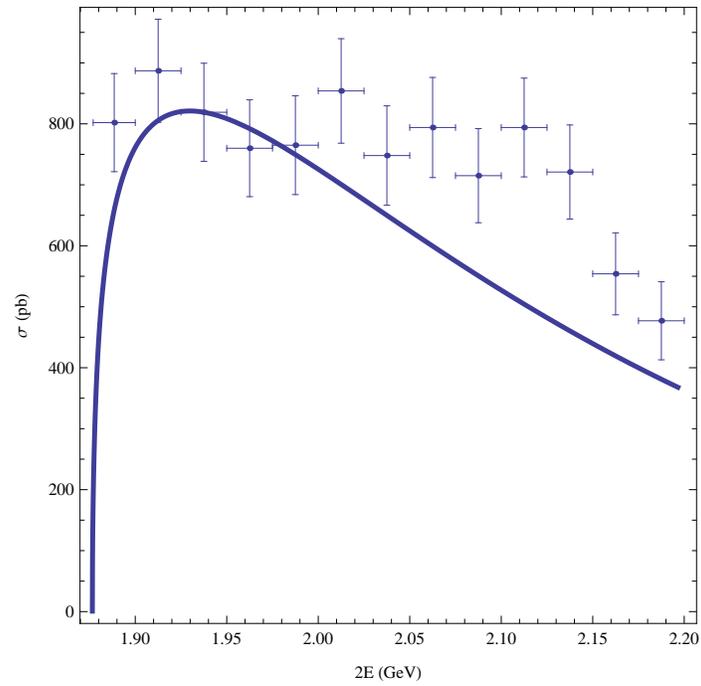}
\caption{Calculated isovector cross section, normalized to the data at the third point.}
\end{figure}
\newpage
%
\begin{figure}
\includegraphics{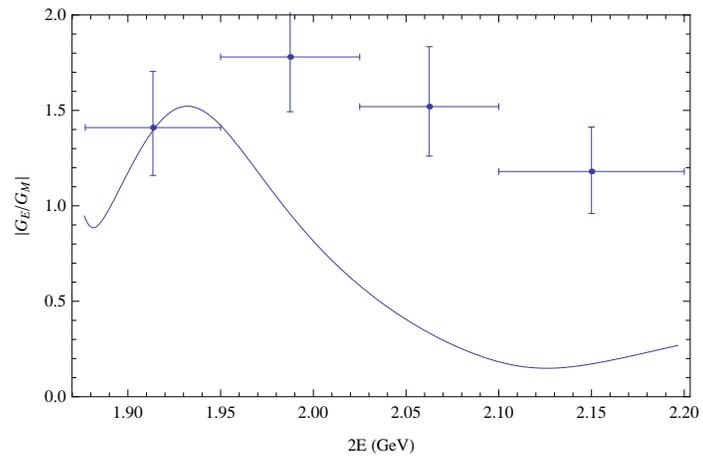}
\caption{Calculated ratio $|G_E^0/G_M^0|$. No free parameters for fixed optical potential. }
\end{figure}
\end{document}